\documentclass{aa}
\usepackage{graphicx}
\usepackage{txfonts}
\usepackage{hyperref}
\hypersetup{colorlinks=true, breaklinks=true, linkcolor=red, citecolor=blue, urlcolor=blue}
\begin{document}

\title{Evolution of the flow field in decaying active regions\thanks{This paper is mainly based on Part\,II of the Ph.D. thesis ``On the decay of sunspots'', \url{https://freidok.uni-freiburg.de/data/165760}.}}
\titlerunning{Evolution of the flow field in decaying active regions}
\subtitle{II. Converging flows at the periphery of naked spots}
\authorrunning{Strecker \& Bello Gonz\'alez}
\author{H. Strecker\inst{1,2} \and N. Bello Gonz\'alez\inst{1}} 
\institute{%
 Leibniz-Institut f{\"u}r Sonnenphysik (KIS),
 Sch{\"o}neckstr.\ 6,
 79104 Freiburg,
 Germany
\and
 Instituto de Astrof\'isica de Andaluc\'ia (CSIC),
 Apdo. de Correos 3004,
 E-18080 Granada,
 Spain\\
 \email{streckerh@iaa.es}
}
\date{Received 2 November 2021 / Accepted 27 April 2022}
\abstract
{In a previous work, we investigated the evolution of the flow field around sunspots during sunspot decay and compared it with the flow field of supergranular cells. The decay of a sunspot proceeds as it interacts with its surroundings. This is manifested by the changes observed in the flow field surrounding the decaying spot.
} 
{We now investigate in detail the evolution of the flow field in the direct periphery of the sunspots of the same sample and aim to provide a complete picture of the role of large-scale flows present in sunspot cells.
}
{We analyse the horizontal velocity profiles of sunspots obtained from observations by the Helioseismic and Magnetic Imager (HMI) on board  the Solar Dynamics Observatory (SDO). We follow their evolution across the solar disc from their stable phase to their decay and their final disappearance.
}
{We find two different scenarios for the evolution of the flow region surrounding a spot in the final stage of its decay: (i) either the flow cell implodes and disappears under the action of the surrounding supergranules or (ii) it outlives the spot. In the later case, an inwards flow towards the remaining naked spot develops in the vicinity closest to the spot followed by an outflow further out. These findings provide observational evidence to theoretical predictions by realistic magnetohydrodynamic (MHD) sunspot and moat region simulations.
}
{The Evershed flow and the moat flow, both connected to the presence of fully fledged sunspots in a spot cell, vanish when penumbrae decay. Moat flows decline into supergranular flows. The final fate of a spot cell depends on its interaction with the surrounding supergranular cells. In the case of non-imploding spot cells, the remaining naked spot develops a converging inflow driven by radiative cooling and a geometrical alignment of granules in its periphery which is similar to that observed in pores.
}
\maketitle
%
\section{Introduction}\label{sec:intro}
We build this analysis on the study of the evolution of the moat flow into a supergranular-like flow which we reported in \citet{Strecker_2018}, hereafter referred to as Paper\,I. The report by \citet{Rempel_2015} of an inflow around a simulated naked spot which is a sunspot that has lost its penumbra motivated us to further analyse the existence of a flow in the direct proximity of naked spots in the final stage of sunspot decay. \citet{Rempel_2015} described the simulated naked spot as surrounded by an inflow with a peak velocity of 2\,km\,s\textsuperscript{-1}. The flow is located at the boundary of the naked spot and extends over a radial distance of less than 2\,Mm from the spot boundary. It is enclosed by an outflow which shows similarities to a supergranular flow.\par
Inflows or converging flows around developing sunspots which have not yet formed a penumbra \citep[see e.g.][]{SainzDalda_2012} or around pores are a well-known phenomenon \citep[see e.g.][]{Wang_1992,Roudier_2002,VargasDominguez_2010}. \citet{Cameron_2007} described horizontal flows towards a pore in 3D magnetohydrodynamic (MHD) simulations and concluded that the flow maintains the magnetic structure of the pore.\par
The main flow pattern related to fully developed, stable sunspots in the photosphere is directed outwards. Within the penumbra, a radial outward-directed flow is measured. This is called the Evershed flow \citep{Evershed_1909}. This outflow is magnetised and largest in regions where the magnetic field is almost horizontal. It increases from the inner to the outer penumbra \citep[e.g.][]{Schlichenmaier_2000}. Peak velocities up to 10\,km\,s$^{-1}$ are reported with average horizontal velocities in the range of 2\,--\,4\,km\,s$^{-1}$ \citep{Shine_1994,Schlichenmaier_2000, Loehner-Boettcher_2013}. The outflow is fastest in the intraspines of the penumbra. Intraspines are nearly horizontal magnetic flow channels where the Evershed flow is confined \citep{LR_Borrero_2011}. In the inner penumbra, upflows are measured while the outer penumbra is dominated by measurable downflows \citep[e.g.][]{Schlichenmaier_2000}.\par
Directly at the periphery of a fully fledged sunspot, a horizontal plasma flow which is directed radially outwards is measured. This outflow leads to the development of an annular cell around the sunspot, the so-called moat, which is mainly free of magnetic flux. The flow was first detected by \citet{Sheeley_1972}. It has an average radial extension of 9\,Mm with values ranging from 5\,Mm up to 20\,Mm from the sunspot boundary \citep[e.g.][]{Brickhouse_1988,Loehner-Boettcher_2013,Verma_2018}. The outflow is mainly unmagnetised. This differentiates it from the magnetised Evershed flow. It should be noted that both flows show no obvious connection despite their common flow direction \citep{Deng_2007,Verma_2012,Balthasar_2013}. The horizontal velocity component of the moat flow is a function of the radial distance to the sunspot. The maximum velocity is measured in the direct proximity of the sunspot with velocities in between 0.8\,km\,s$^{-1}$ up to 1.4\,km\,s$^{-1}$ \citep{Muller_1987,Balthasar_2013,Loehner-Boettcher_2013}. The maximum velocity decreases almost continuously with increasing radial distance from the sunspot. \citet{Rempel_2015} calculated realistic numerical MHD simulations of a sunspot and its moat region in the photosphere and upper convection zone. The characteristics of the moat flow at the solar surface (at $\tau=1$) agree with characteristics known from observations. Beneath the sunspot penumbra, an upflow is measured. At the boundary of the penumbra at the surface, it changes direction and becomes an outflow. Based on the results of the simulations, the moat flow is caused by a reduction of down flows in the proximity of the sunspot. The up- and downflow balance, which is present within quiet Sun regions, is perturbed. The reduction of these cool downflows leads to a rise in the average temperature in the convection zone in the proximity of the sunspot. This average rise in temperature drives the moat flow \citep{Rempel_2015}.\par
The disappearance of the penumbra influences the surrounding flow region (\citealt{Deng_2007}; Paper\,I). The horizontal flow profile characteristic for the moat flow disappears with the loss of the penumbra (see Paper\,I). \citet{Deng_2007} studied the evolution of one decaying spot and found proper motions towards it in an annular region which separate the outflow region from the spot which had lost its penumbra. They state that the newly evolved flows have much weaker horizontal velocities compared to the original moat flow with the outflow being less stable than the moat flow. A similar evolution is found by \citet{Rempel_2015} in data from a MHD simulation of a naked spot. An inflow of up to 1\,km\,s$^{-1}$ separates the remnant outflow from the sunspot. The outflow has a maximum velocity around 0.5\,km\,s$^{-1}$ at approximately 4\,Mm from the spot boundary. \citet{Verma_2018} describe stable sunspots as part of a large magnetic flux system that are already supergranular cells on their own in this stage. Paper\,I describes the transition of the moat into a supergranular-like flow\footnote{Supergranulation generates a cellular flow pattern which is only visible in Doppler maps. The cells are around 35 times larger than granules. The flow transports magnetic flux to its boundaries and might therefore be correlated with the formation of the magnetic network \citep[see e.g.][]{Simon_1964, Roudier_2014,OrozcoSuarez_2012}} when the penumbra dissolves. When the penumbra has dissolved, the maximum horizontal velocity is no longer localised at the sunspot boundary. Instead, the horizontal velocity increases with radial distance from the spot, reaches a maximum velocity within the flow region, and decreases further out towards the network. Paper\,I conjectures that surrounding supergranules might disturb the annular flow pattern. The weaker magnetic flux of the naked spot is advected and part of it can be swept towards the network. Thus, the magnetic field fragments. A significant amount of magnetic flux therein gets lost due to the cancellation of flux of opposite polarity, for example, at the network \citep{LR_Driel_2015}. The disappearance of the naked spot in intensity maps is followed by two possible scenarios for the remnant region. (1) The surrounding supergranules squeeze the magnetic flux and it becomes part of the network. (2) The flow cell can persist and the magnetic flux would then be transported towards the network.\par
We make use of the same data set which we analysed in Paper\,I and focus our analysis on the horizontal flow in the region within the sunspot and its closest proximity. This region was excluded in Paper\,I where the flow was analysed in regions $r_\text{F}\ge$\,3\,pix\,=\,1.1\,Mm (with $r_\text{F}$\,=\,0\,Mm at the boundary of the sunspot). Section\,\ref{sec:data} describes the analysed data and the analysis method we used to obtain horizontal flow components from Doppler maps. In Sect.\,\ref{sec:results}, we present the results for the evolution of the horizontal flow properties within and at the periphery of the decaying spots. We discuss the results in Sect.\,\ref{sec:disc} by taking the evolution of the flow region at a larger distance from the sunspot into account. This leads to a complete picture of the evolution of the radial flow profile related to decaying spots which we describe in Sect.\,\ref{sec:conc}.
\section{Data and methods}\label{sec:data}
\begin{figure*}[ht!]
\begin{center}
\includegraphics[width=1.\textwidth]{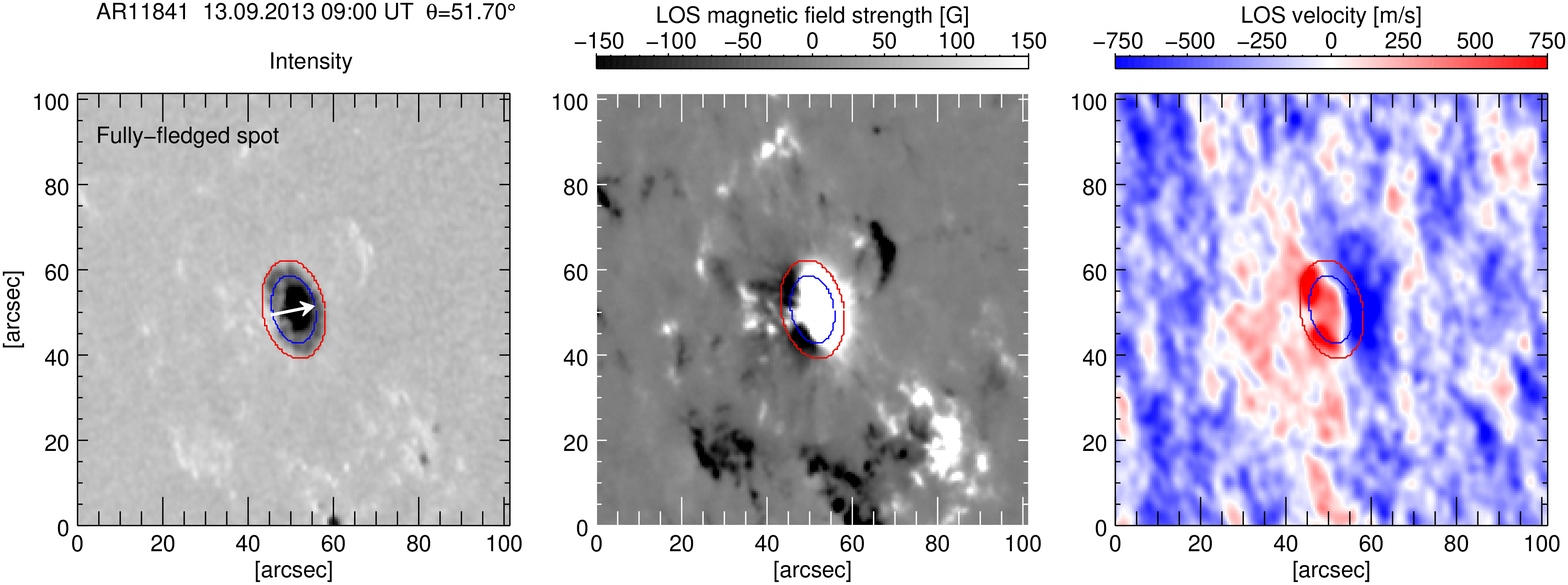}

\includegraphics[width=1.\textwidth]{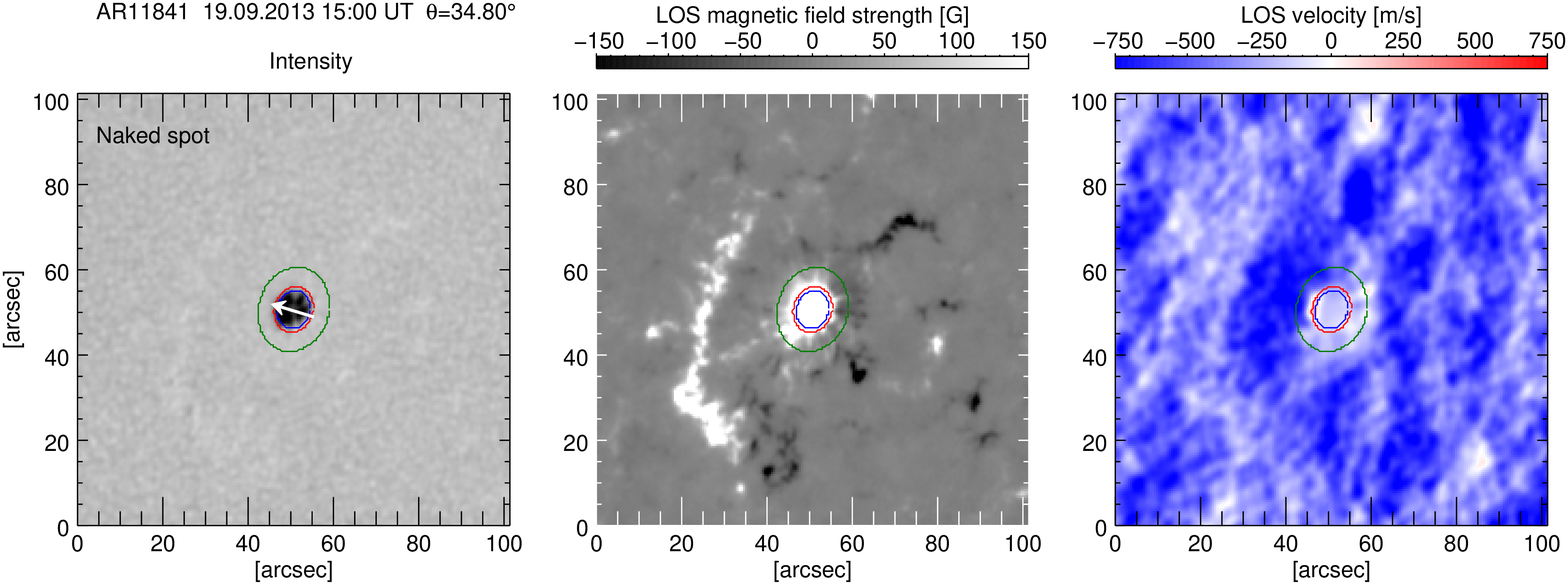}
\end{center}
\caption{Intensity map (left), LOS magnetogram (middle), and Doppler map (right) show the sunspot of AR11841 on the first day (top) of the analysis and six days later (bottom). The position of the maximum horizontal velocity within the spot, within the surroundings, and the radius of the spot are marked by blue, green, and red ellipses, respectively. The arrow in the intensity map points in the direction of disc centre.\label{fig:imd_maps}}
\end{figure*}

In this paper, we make use of a data set we selected in Paper\,I to study the evolution of the moat flow into a supergranular flow during sunspot decay. The data set consists of eight sunspots. The sunspots are roundish and fully fledged when they are localised close to the eastern limb, that is, at the beginning of the analysis. Their decay process can be observed while they move towards the western limb. The data set is based on data from the Helioseismic and Magnetic Imager \citep[HMI, ][]{Schou_2012} on board the Solar Dynamics Observatory \citep[SDO, ][]{Pesnell_2012}. HMI 720\,s Doppler maps are used to study the flows within and in the regions surrounding the spots. In addition, we use intensity maps to localise the spots and line-of-sight (LOS) magnetograms as context data. For all three data products, we generated 3\,h time averages (out of 15 successive Doppler maps). The averaging has led to a total of eight data sets per day. The sunspot was localised in intensity maps (see Fig.\,\ref{fig:imd_maps}, left panels) and its outline was obtained by an intensity threshold $I_{c}$\,=\,0.9\,$I_{qs}$ with the normalised intensity of the quiet Sun, $I_{qs}$. We assumed the spot was circular so we defined the boundary of the spot as the maximum distance between the contour line and the centre of gravity of the spot (see Fig.\,\ref{fig:imd_maps}, red ellipses). We determined the time of disappearance of the penumbra by eye while we defined the time of disappearance of the naked spot using the automatic localization of the spot in intensity maps. For a detailed description of the reduction of the Doppler maps, the selection of the eight analysed sunspots, and the localisation of the spots within the maps, we refer the reader to Sect.\,2 of Paper\,I.
%
\paragraph{Analysis of the flows of spots.}\label{sec:ana}
The analysis method is based on the work by \citet{Loehner-Boettcher_2013}, Paper\,I, and \citet{Strecker_2019} and focusses on the analysis of horizontal flows in Doppler maps. For the analysis, we assume the sunspots are circular with axially symmetrical flows. Then azimuthally averaged flow properties can be determined \citep{Schlichenmaier_2000}. The LOS velocity is read out along circles with different radii in steps of one pixel around the centre of the sunspot. The LOS horizontal velocity component, $v^{\text{LOS}}_{\text{h}}(r,\theta)$ is obtained as the amplitude from a sine fit along the circles. We make use of the heliocentric angle to calculate the horizontal velocity $v_{\text{h}}(r)$ which only depends on the radius, $r$, the distance to the centre of the spot (see Eq.\,1 in Paper\,I). Thus, a radial profile of the horizontal velocity of the flow system of a spot can be determined.\par
\begin{figure*}[ht!] 
\begin{minipage}[r]{0.33\textwidth}{\includegraphics[width=1.\linewidth]{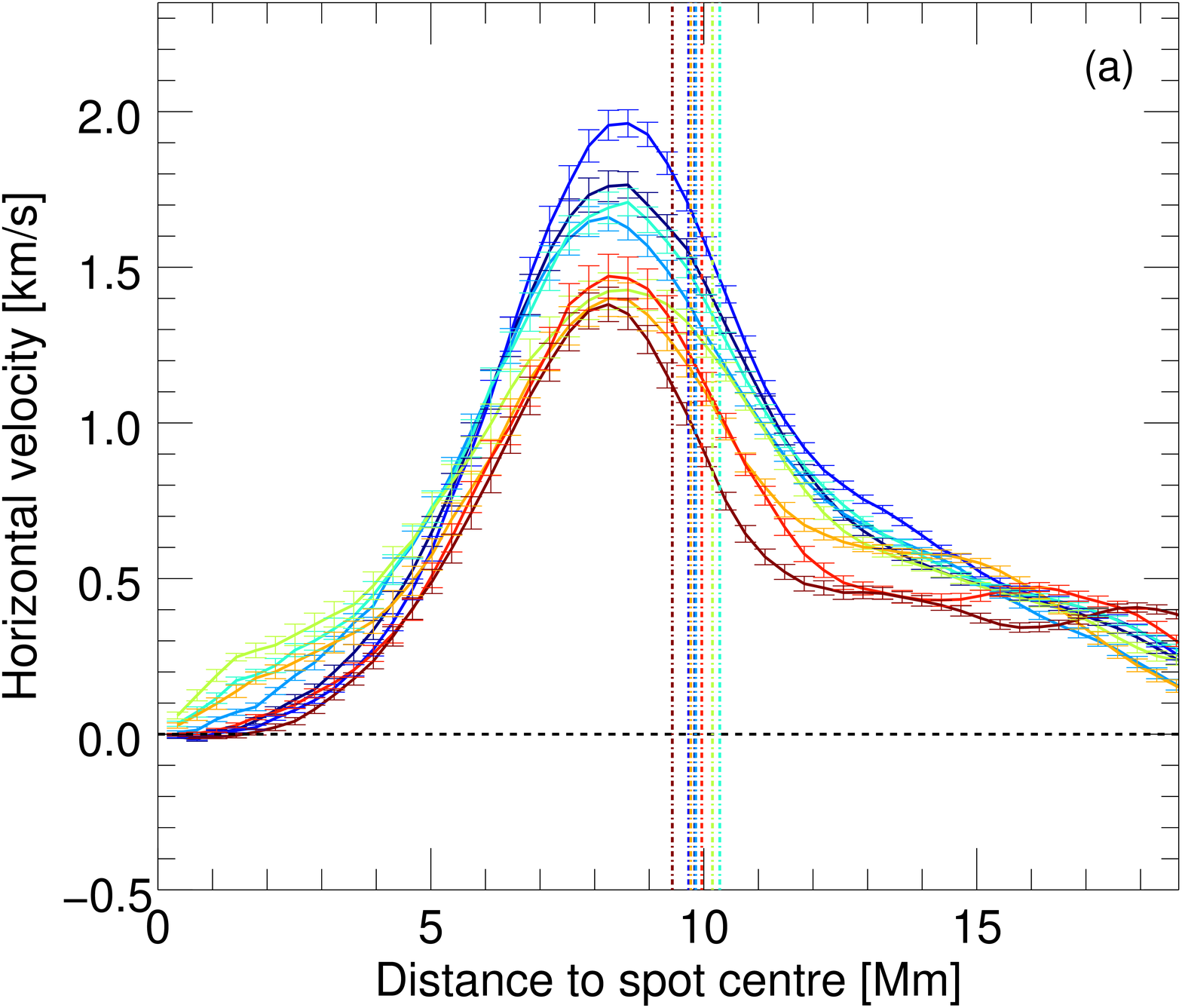}}
\end{minipage}
\begin{minipage}[l]{0.33\textwidth}{\includegraphics[width=1.\linewidth]{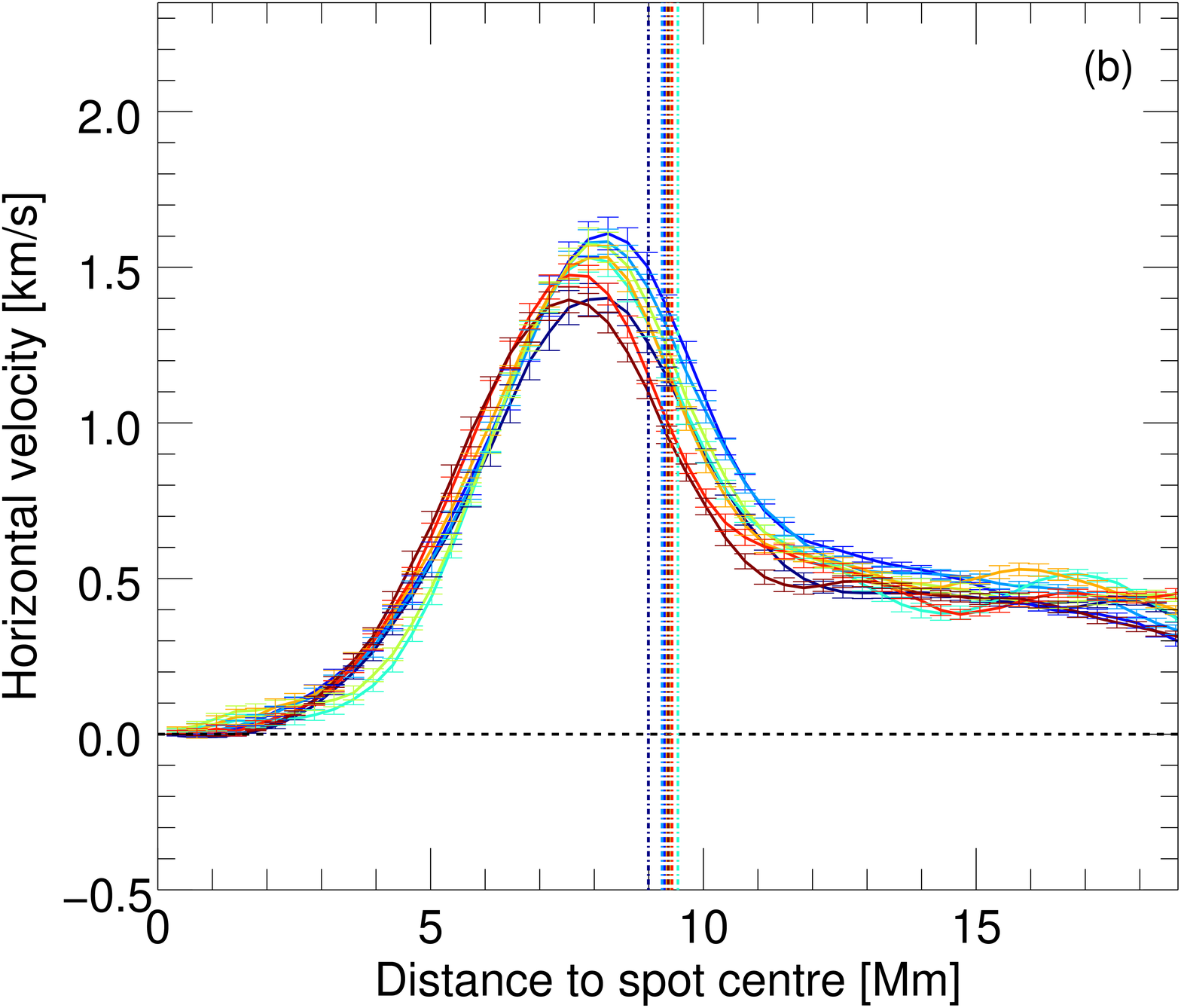}}
\end{minipage}
\begin{minipage}[c]{0.33\textwidth}{\includegraphics[width=1.\linewidth]{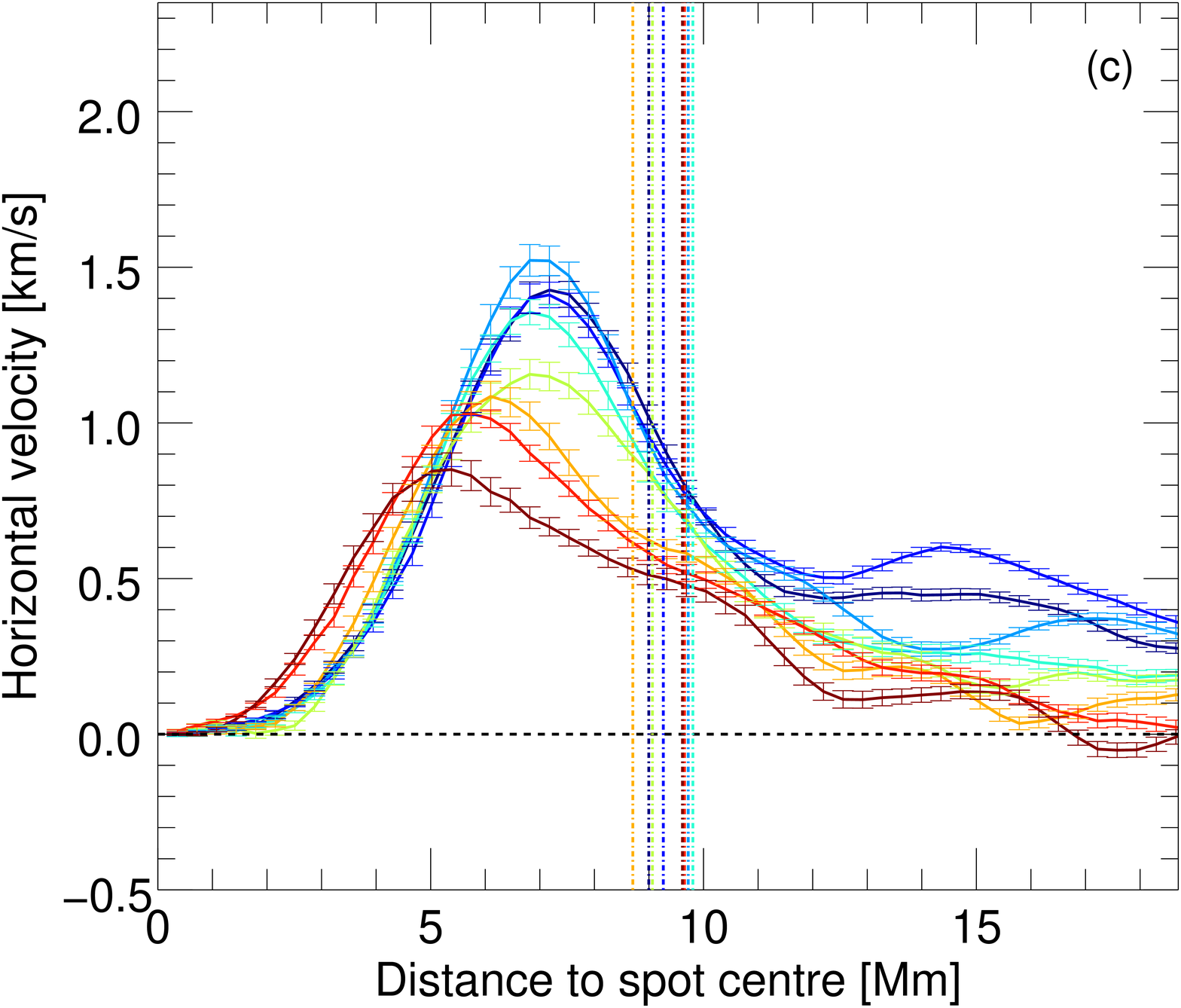}}
\end{minipage} \\
\begin{minipage}[r]{0.33\textwidth}{\includegraphics[width=1.\linewidth]{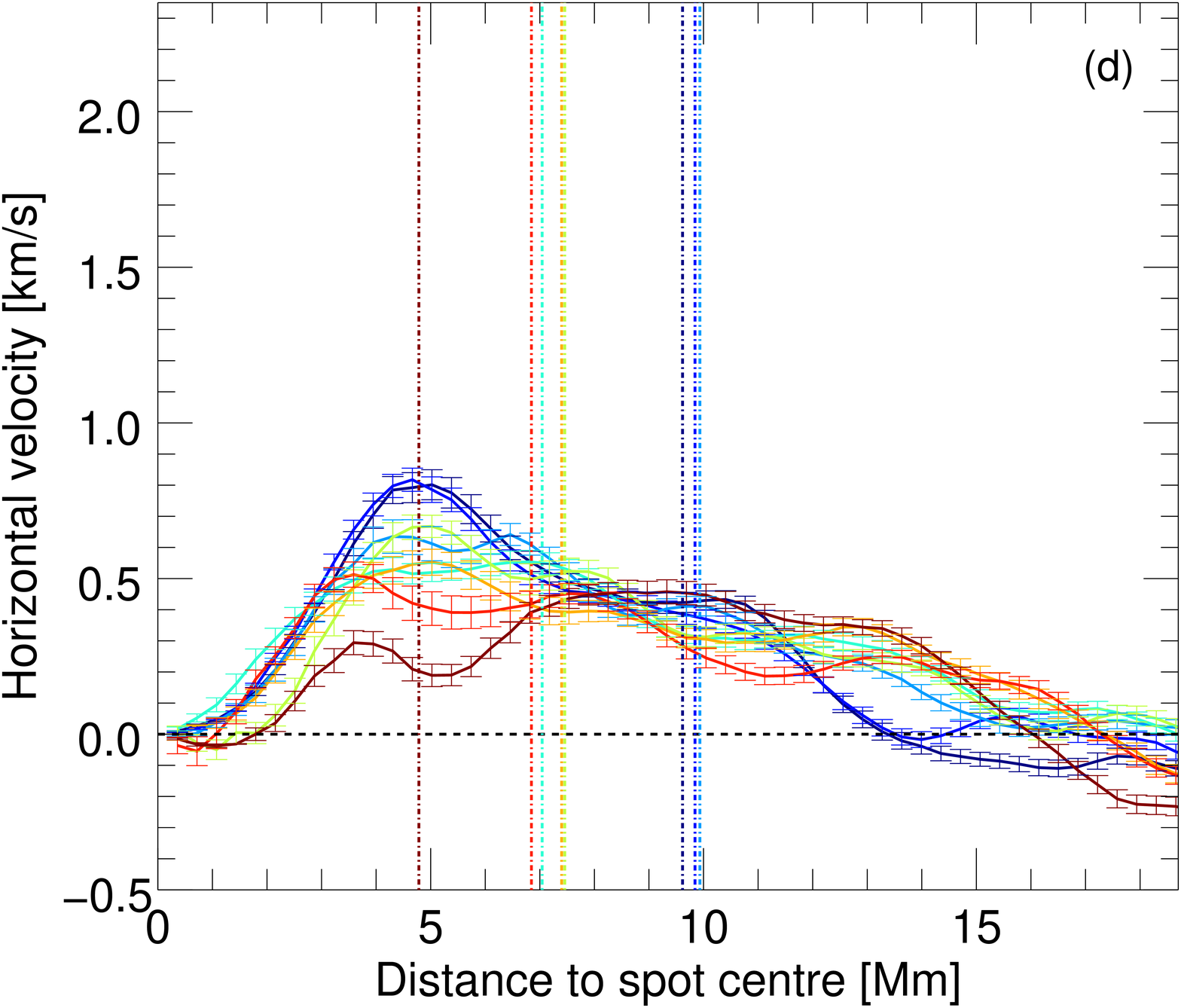}}
\end{minipage}
\begin{minipage}[l]{0.33\textwidth}{\includegraphics[width=1.\linewidth]{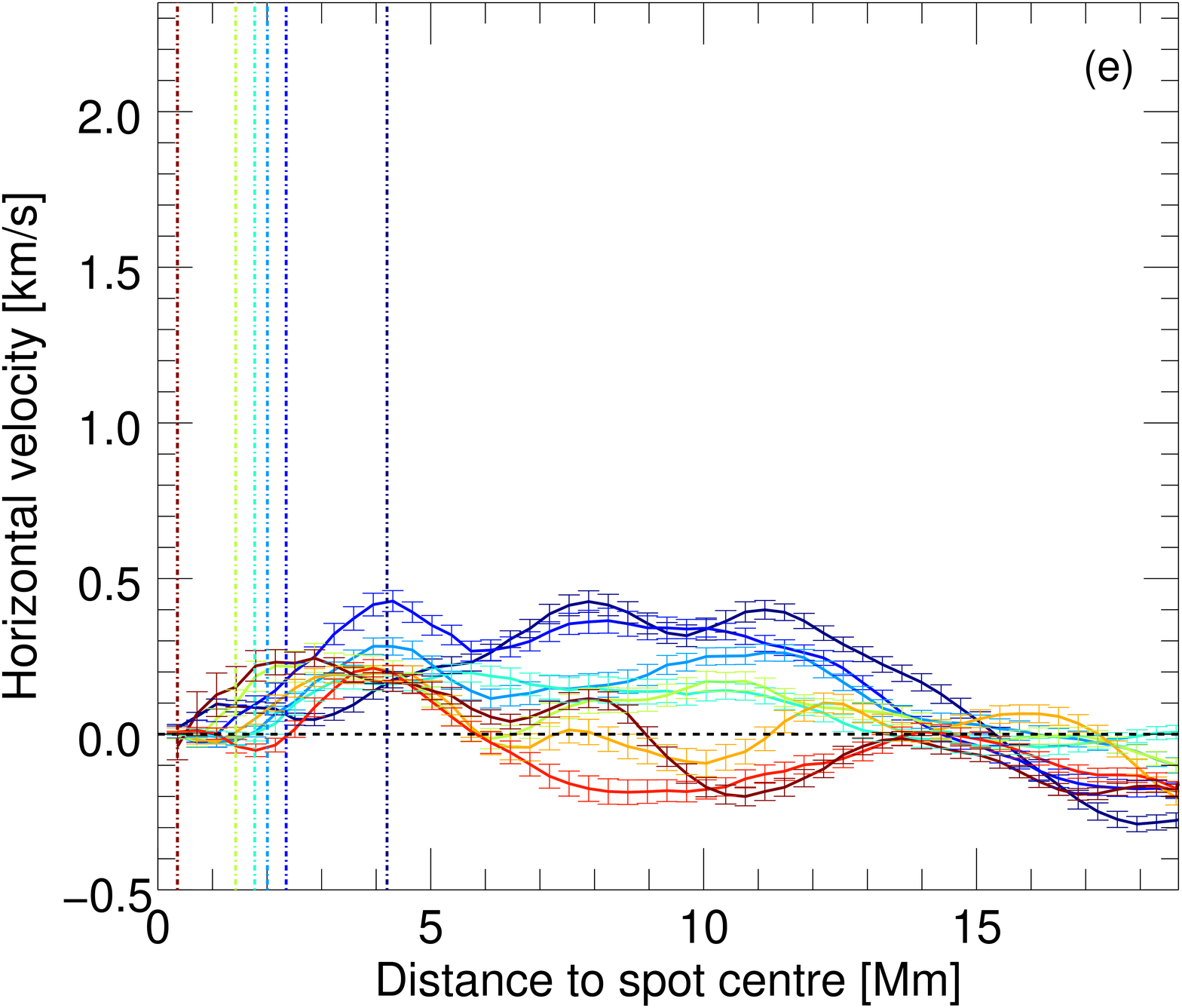}}
\end{minipage}
\begin{minipage}[c]{0.33\textwidth}{\includegraphics[width=1.\linewidth]{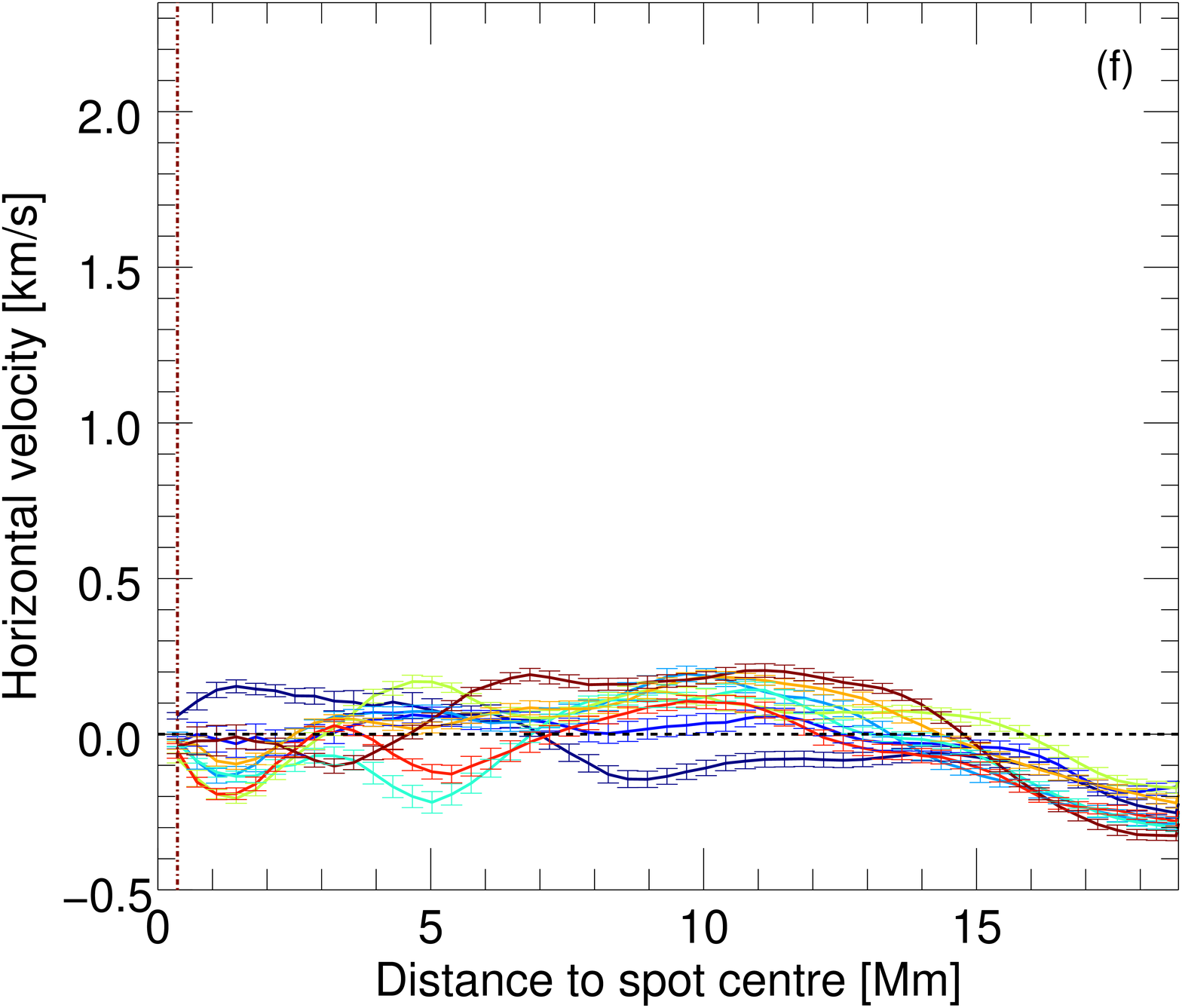}}
\end{minipage}
\caption{Radial profiles of the horizontal velocity of the flow system of AR11646 for six successive days (panel (a) to (f)), representative for Case (i) of the evolution of the flow field. Eight time steps are equally distributed across one day (colour-coding from blue, meaning early, to red, indicating late). The penumbra has dissolved on the the fourth day of the analysis at time step 4 (turquoise line in panel (d)). Error bars represent the standard deviation. Vertical lines display the position of the spot boundary.
\label{fig:inflow_AR11646}}
\end{figure*}
\begin{figure*}[ht!]
\begin{minipage}[l]{0.33\textwidth}{\includegraphics[width=1.\linewidth]{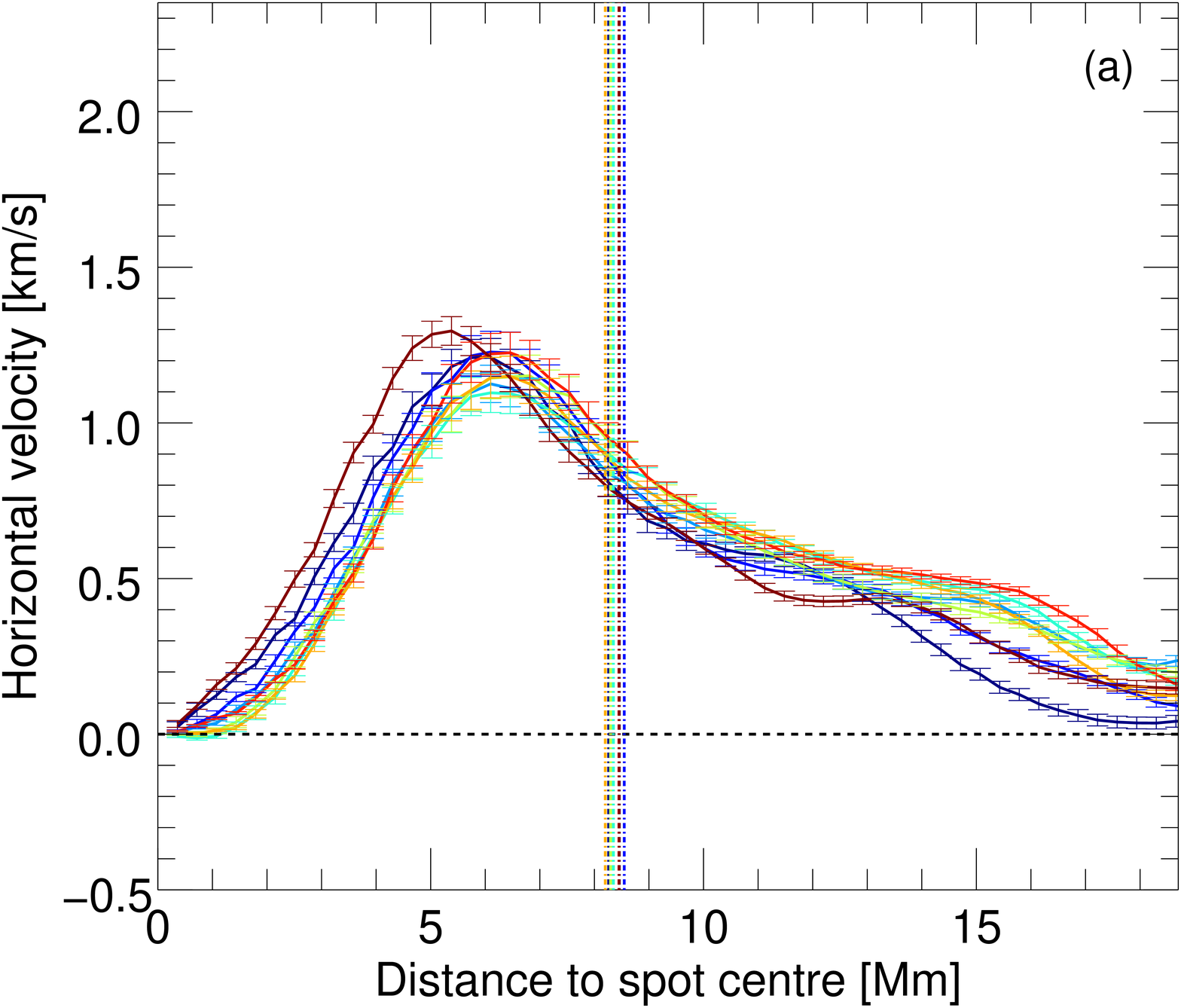}}
\end{minipage}
\begin{minipage}[c]{0.33\textwidth}{\includegraphics[width=1.\linewidth]{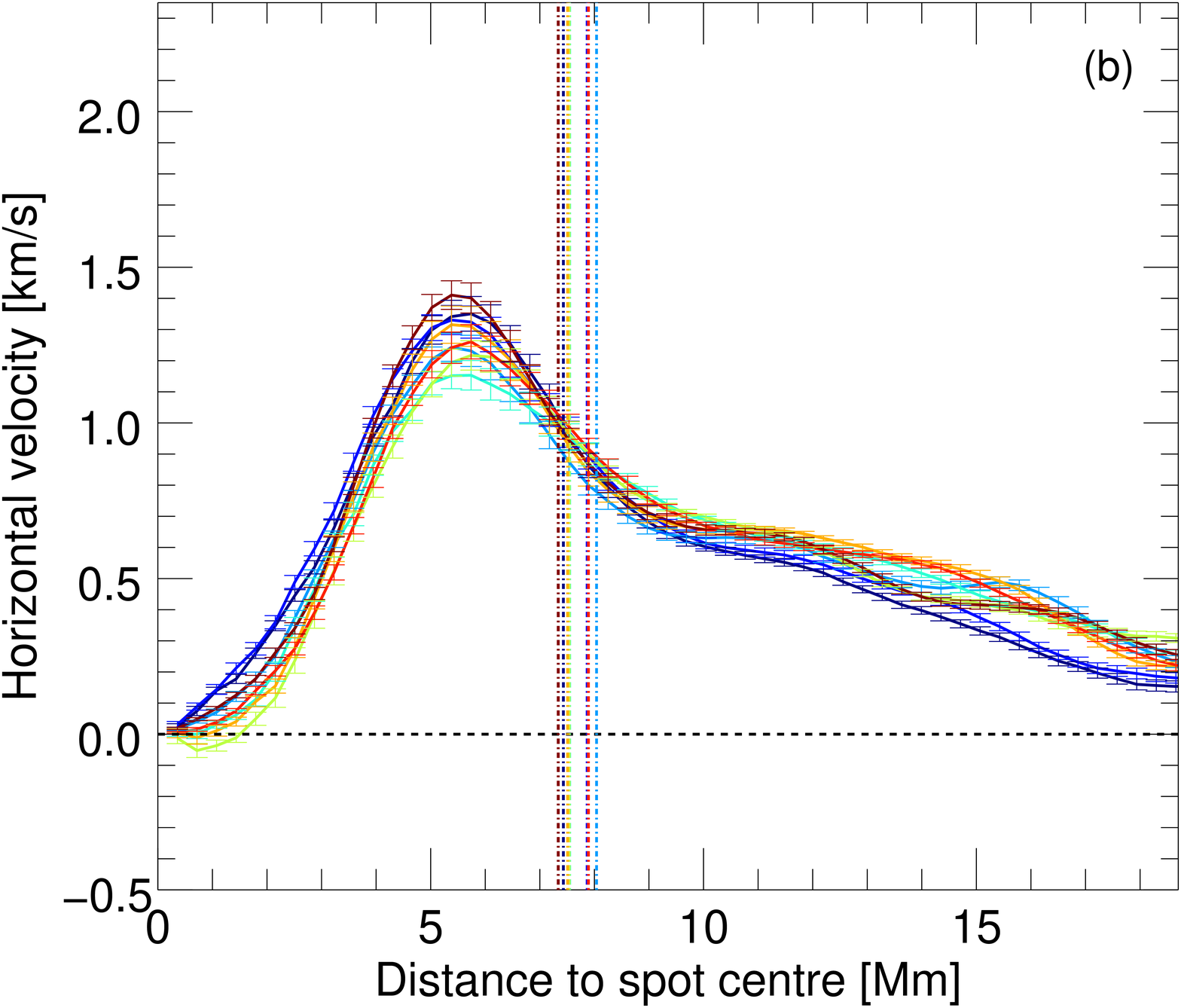}} 
\end{minipage} 
\begin{minipage}[r]{0.33\textwidth}{\includegraphics[width=1.\linewidth]{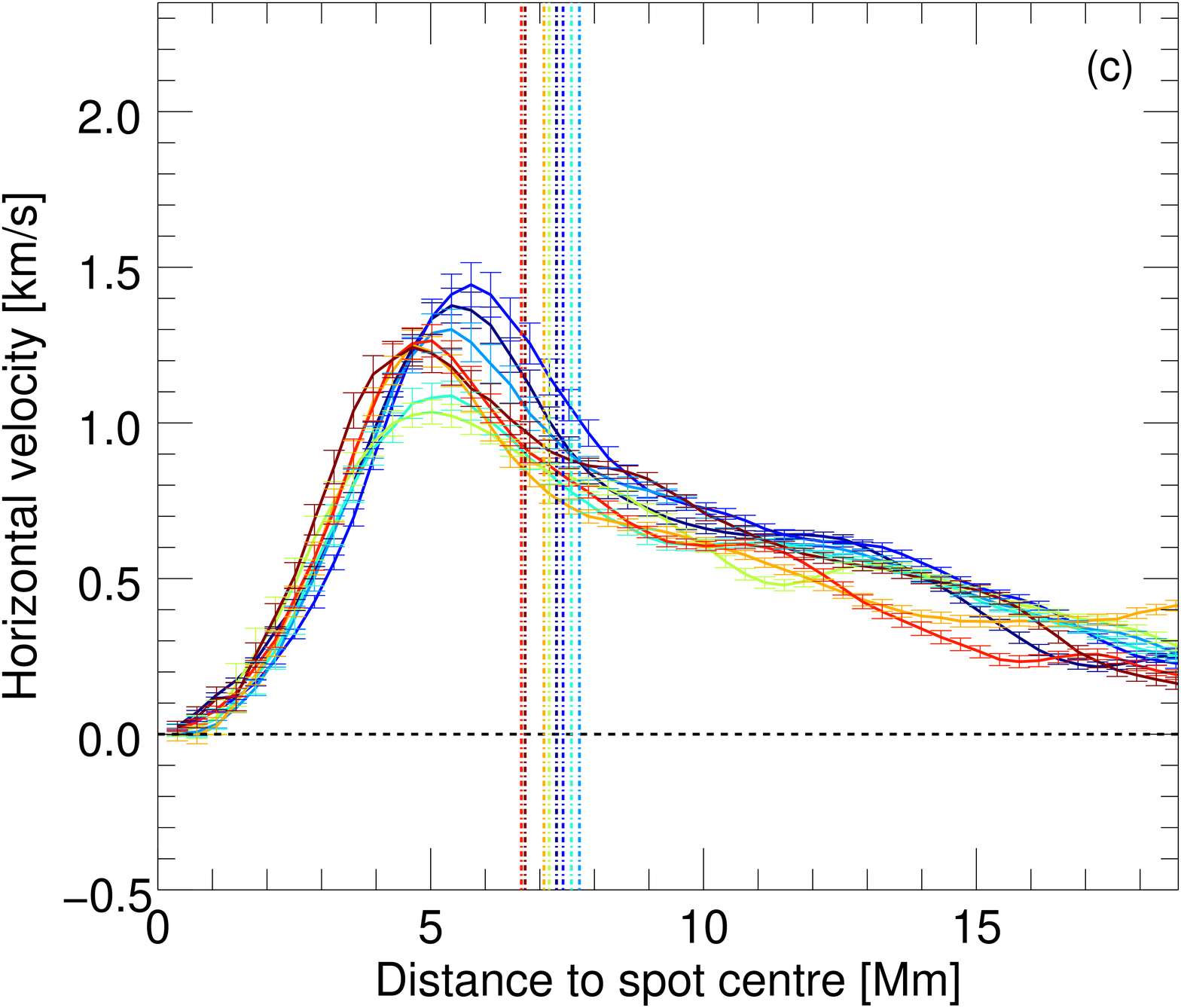}} 
\end{minipage}\\
\begin{minipage}[l]{0.33\textwidth}{\includegraphics[width=1.\linewidth]{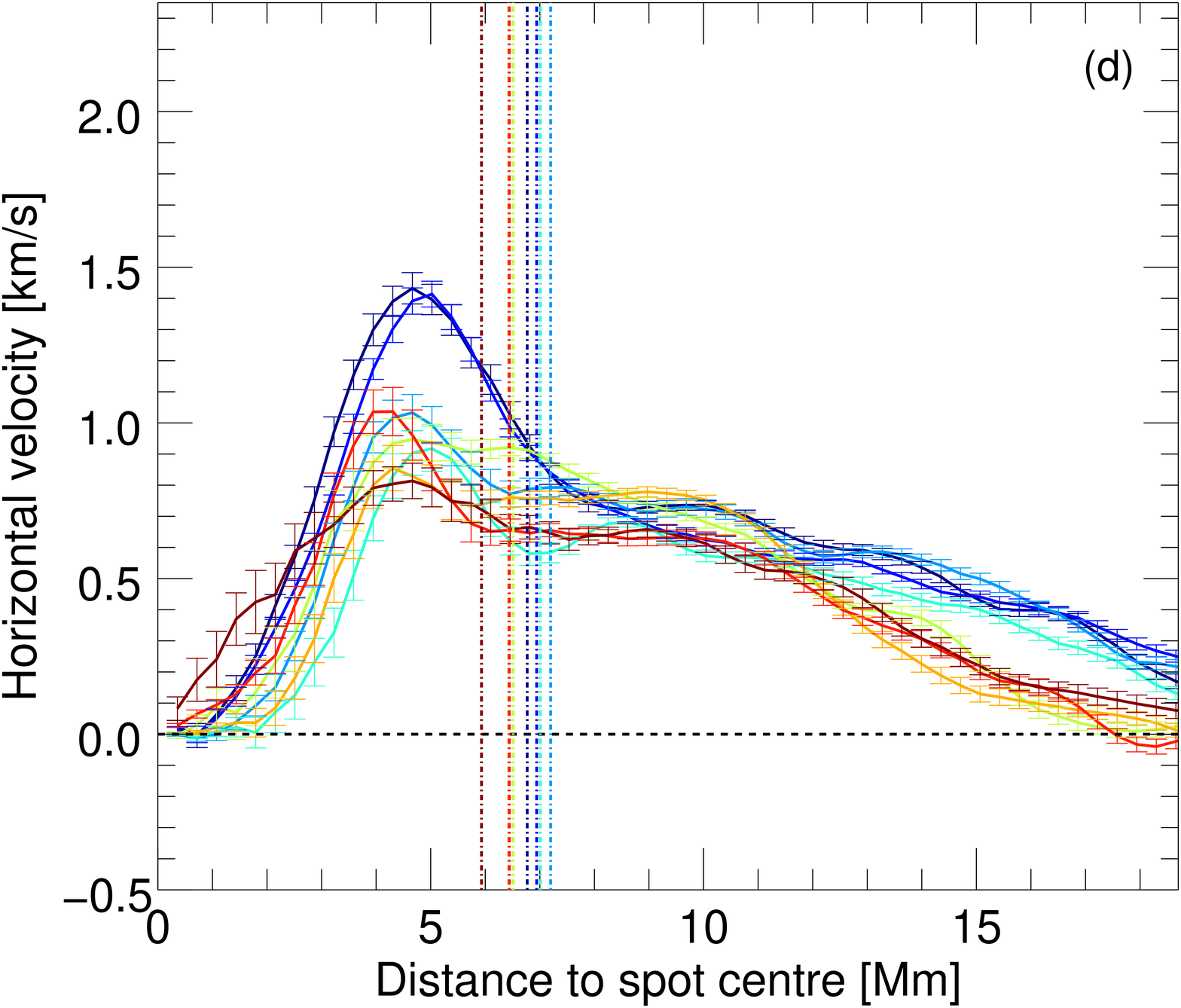}}
\end{minipage}
\begin{minipage}[c]{0.33\textwidth}{\includegraphics[width=1.\linewidth]{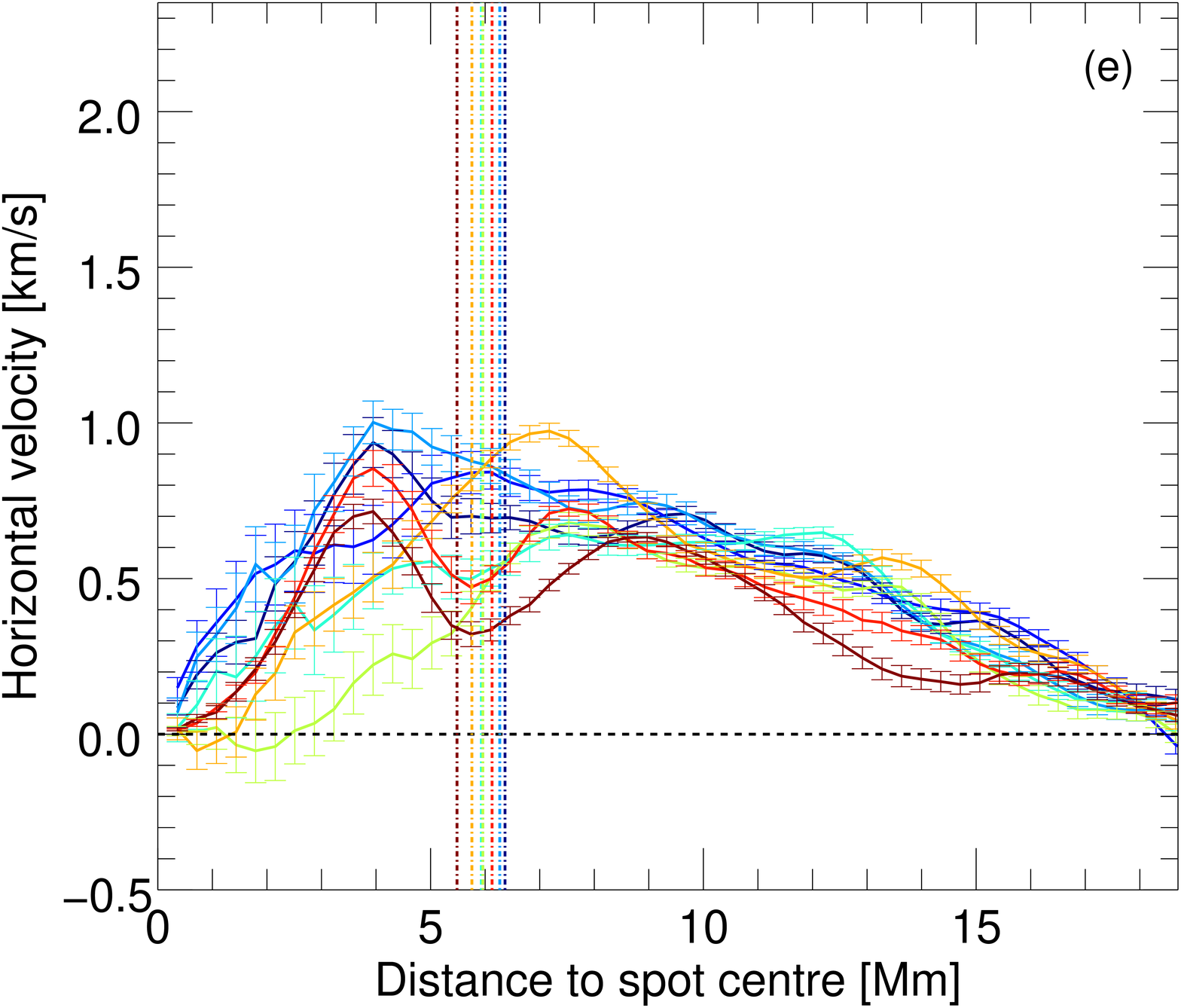}}
\end{minipage} 
\begin{minipage}[r]{0.33\textwidth}{\includegraphics[width=1.\linewidth]{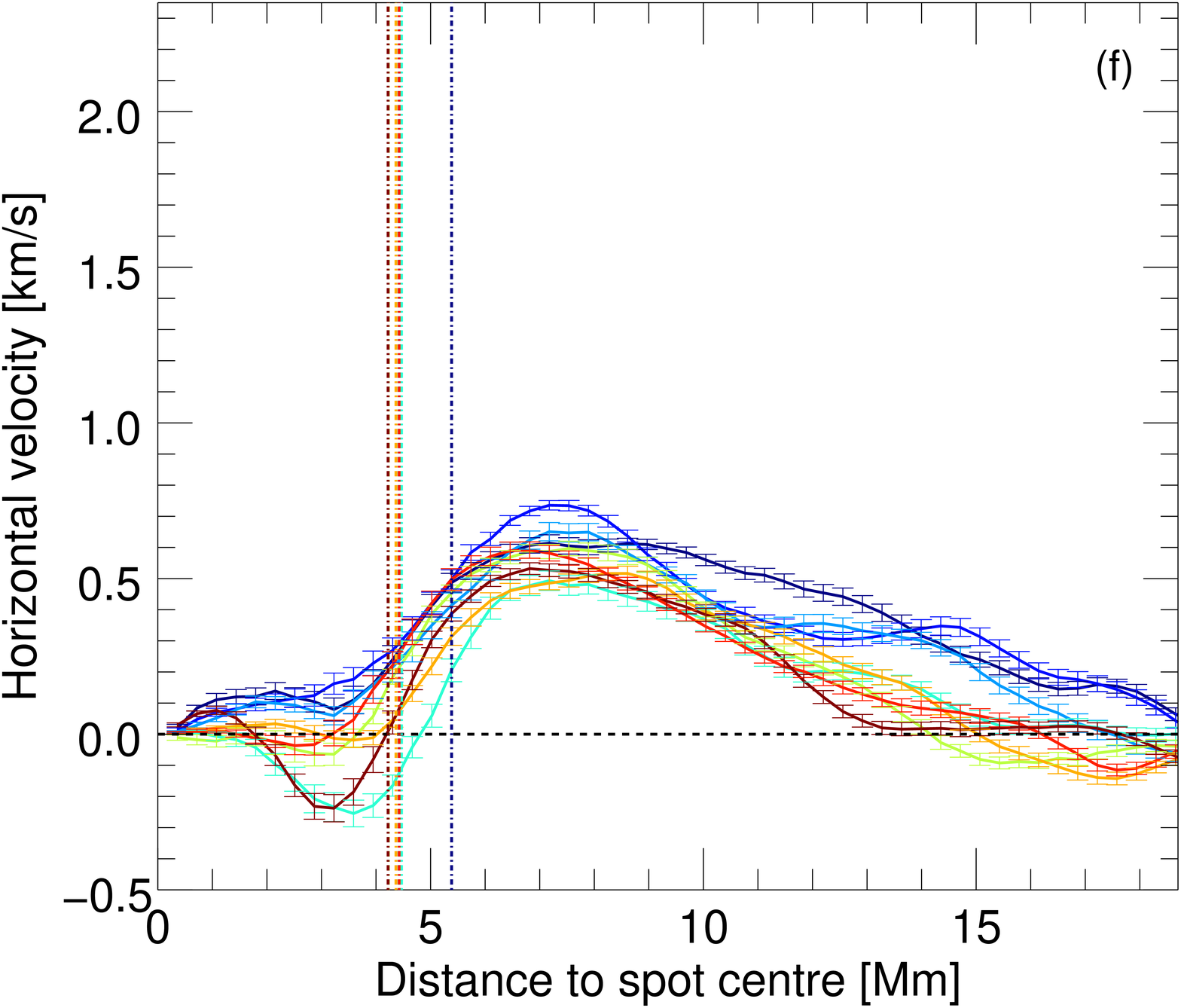}} 
\end{minipage} \\
\begin{minipage}[l]{0.33\textwidth}{\includegraphics[width=1.\linewidth]{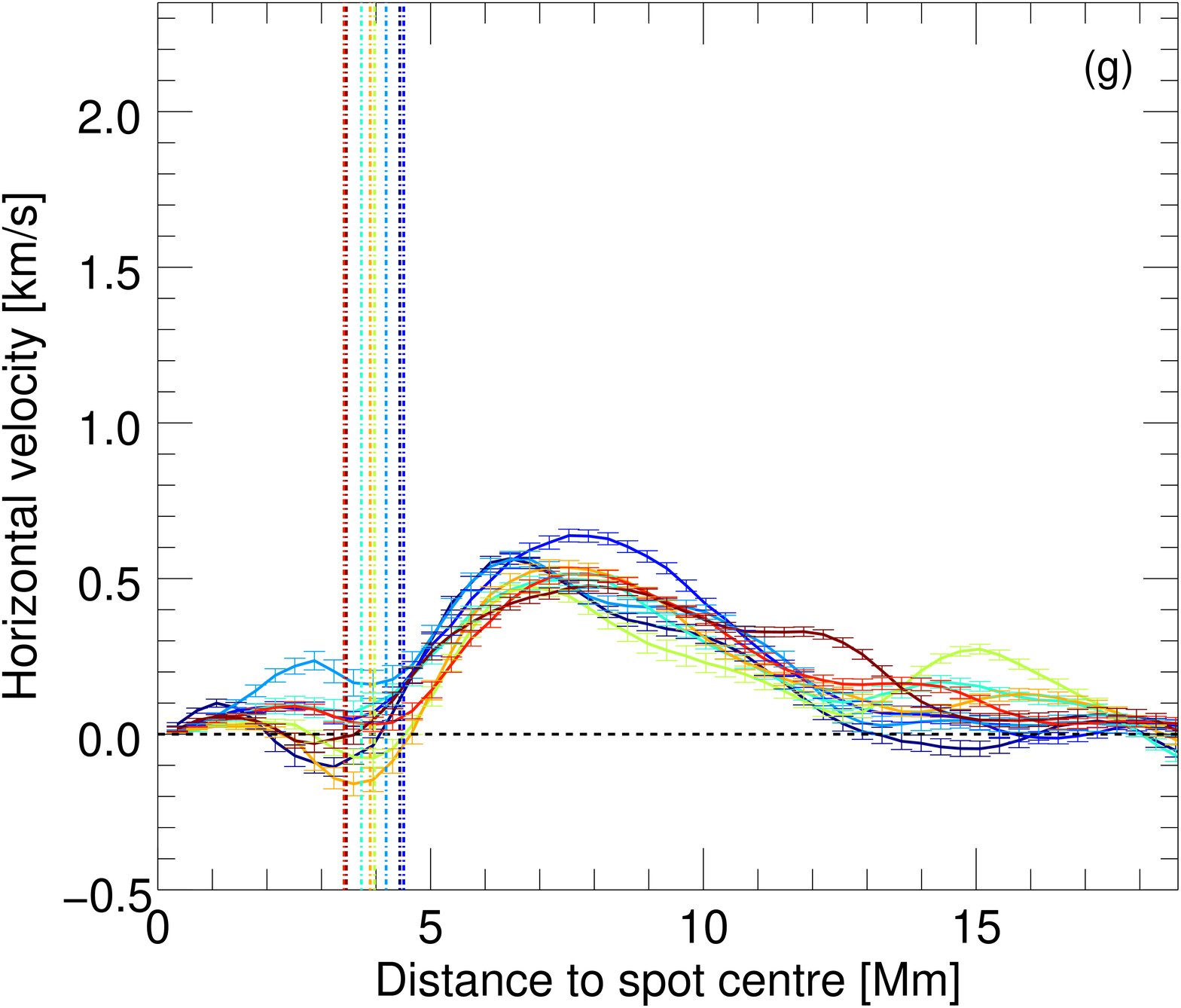}}
\end{minipage}
\begin{minipage}[c]{0.33\textwidth}{\includegraphics[width=1.\linewidth]{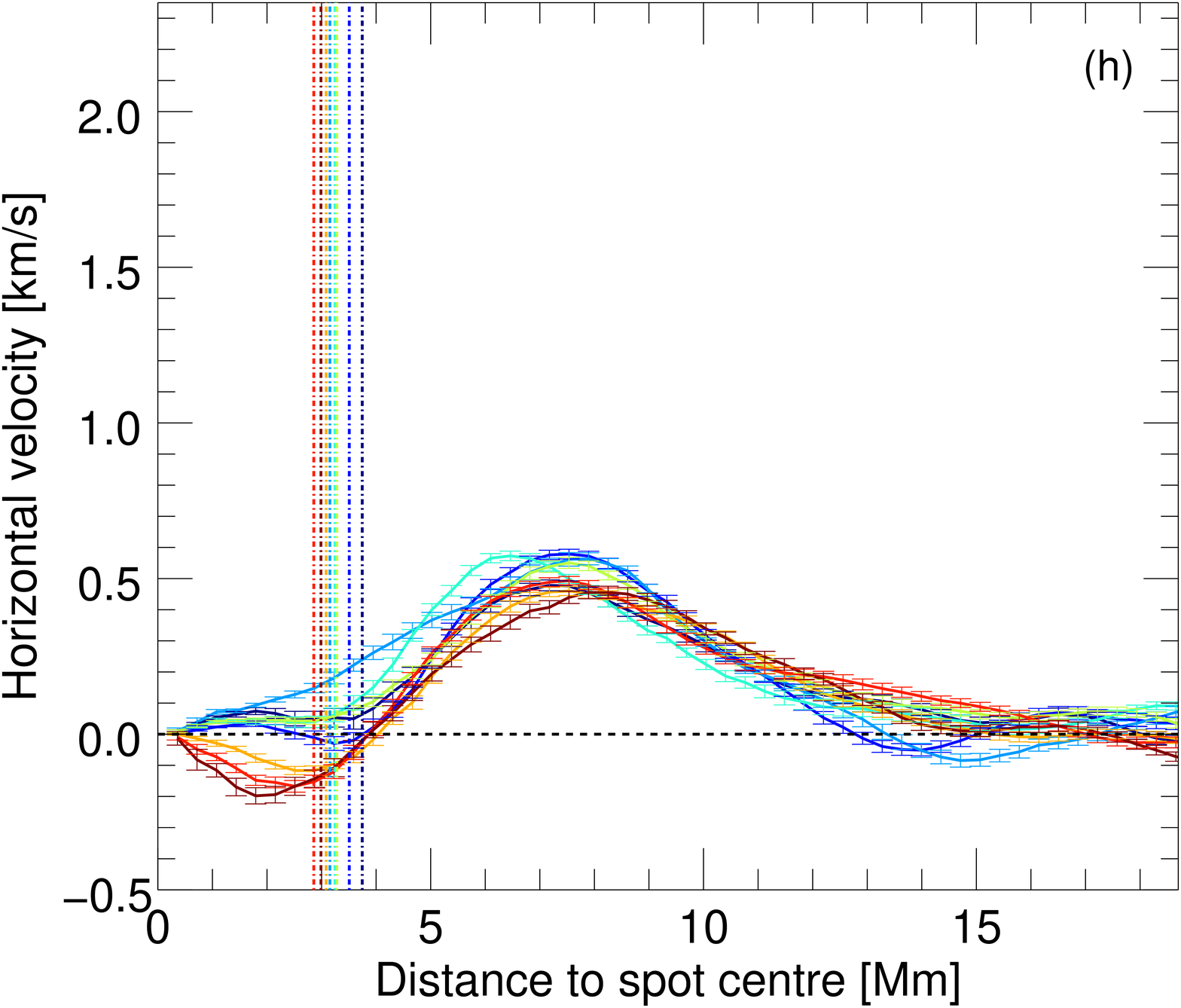}} 
\end{minipage} 
\begin{minipage}[r]{0.33\textwidth}{\includegraphics[width=1.\linewidth]{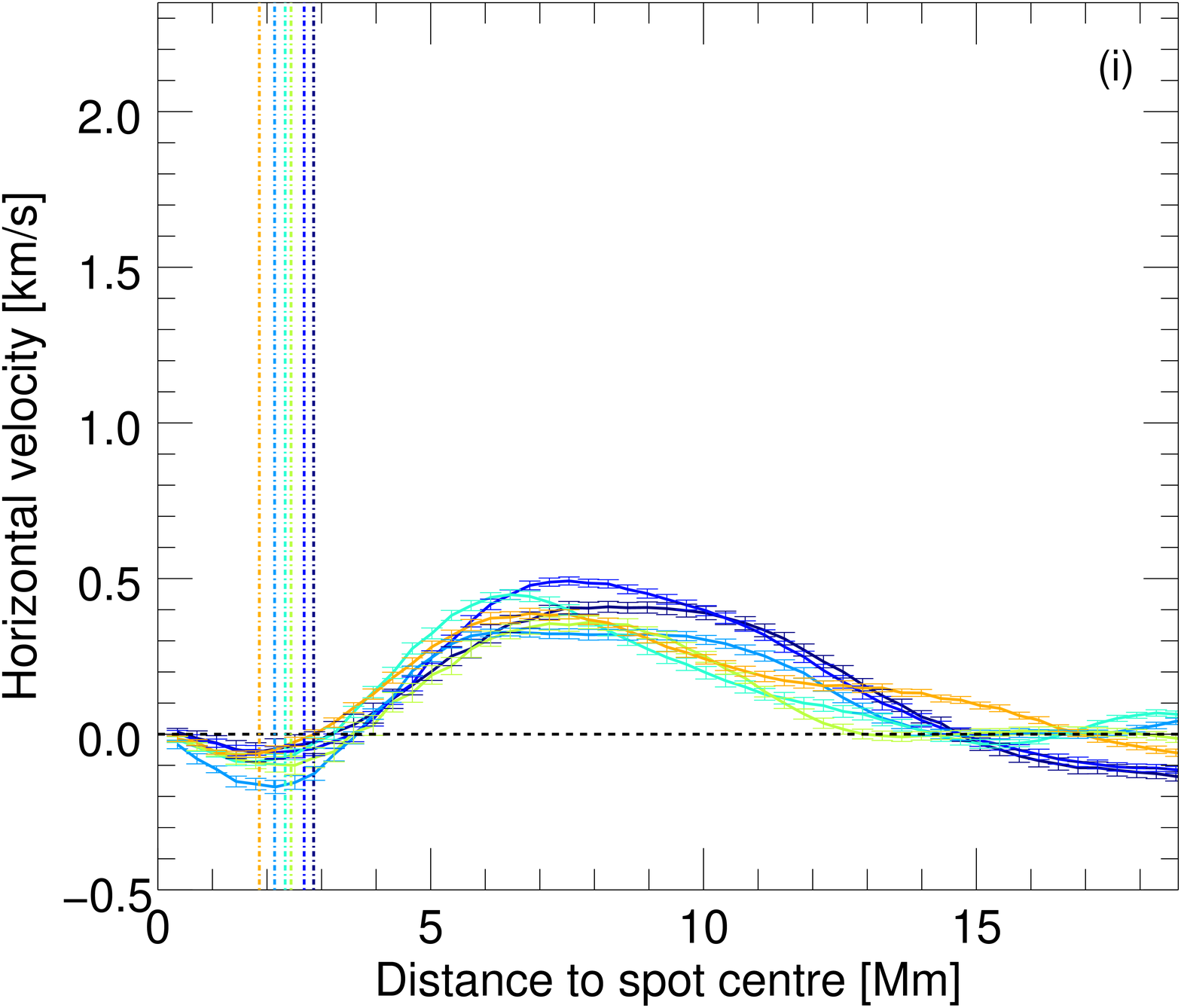}}
\end{minipage}
\caption{Same as Fig.\,\ref{fig:inflow_AR11646}, but for nine successive days (panel (a) to (i)) of AR11841, representative for Case (ii) of the evolution of the flow field. The penumbra has dissolved on the fourth day of the analysis at time step 5 (green profile in panel (d)). Error bars represent the standard deviation. Vertical lines display the position of the spot boundary.
\label{fig:inflow_AR11841}}
\end{figure*}

A fully developed sunspot is known to host two different flows: the Evershed flow within the penumbra and the moat flow in the surroundings. In time, this flow pattern might change and different flows might develop. We refer to all the flows studied along the radial distance from the spot's centre outwards towards the network as the `flow system' of a spot. The flow profiles can be compared at different stages of the evolution of a spot (see e.g. Figs.\,\ref{fig:inflow_AR11646} and \ref{fig:inflow_AR11841}).\par
%
The rms value for each radial position from the spot's centre describes the azimuthal velocity fluctuation of the extracted velocity values, along the sine fit (see Eq.\,2 in Paper\,I). \citet{Loehner-Boettcher_2013} studied the flow within and in the surroundings of stable sunspots while Paper\,I and \citet{Strecker_2019} studied the evolution of the horizontal flow in the surroundings of sunspots for distances larger than 1.1\,Mm from the sunspot boundary during sunspot decay. Both used the rms-value to determine the outer boundary of the flow surrounding the sunspot. Here, we extend the study of the horizontal flow evolution during sunspot decay to regions within the sunspot and at its direct periphery (which were excluded in Paper\,I). The rms value is again used to determine the validity of small velocities in the final stage of the spots' evolution. Horizontal velocities which are smaller than the obtained rms values are considered to be part of the noise.
\section{Results}\label{sec:results}
We started our analysis when all sunspots were in the stable stage. They consist of an umbra which is completely enclosed by a penumbra. Figures\,\ref{fig:inflow_AR11646} and \ref{fig:inflow_AR11841} illustrate the radial profiles of the horizontal velocity of two of the sunspots of the sample -- AR11646 and AR11841, respectively -- for several successive days (different panels). In addition, we provide the horizontal velocity profile for each time step animated as online material for all analysed active regions. Dashed vertical lines indicate the radius of the sunspot and the heliocentric angle, $\theta$, which provides information of the position of the active region (AR) on the solar disc. Panel (a) of Figs.\,\ref{fig:inflow_AR11646} and \ref{fig:inflow_AR11841} show the profiles for eight successive time steps which represent the radial velocities over the course of a day during the stable phase of the spot. The dashed vertical lines represent the radius of the sunspot boundary. As we described above, the radius was obtained as an approximation as the sunspot is not circular. Therefore, the maximum distance between the centre and the boundary of the sunspot was determined and defined as the radius of the sunspot. We determined the boundary of the sunspot by an intensity threshold as we did in Paper\,I. A common shape in the radial profiles of the horizontal velocity can be found for all sunspots at this stage consisting of: (i) a continuous increase in the horizontal velocity from the centre of the sunspot outwards, (ii) a decrease in the horizontal velocity outwards still within the boundary of the sunspot, and (iii) a further decrease in the horizontal velocity outside the sunspot. However, the velocity profiles flatten as the sunspots decay. The peak velocities obtained for the eight sunspots on the first day of their analysis range from 1.1\,km\,s\textsuperscript{-1} to 2.8\,km\,s\textsuperscript{-1}. Outside the sunspot, at a distance of 3\,pixels from the sunspot boundary, we measured peak velocities in the range from 1.1\,km\,s\textsuperscript{-1} to 0.7\,km\,s\textsuperscript{-1}. Further away from the sunspot, the velocity decreases down to 0.2\,m\,s\textsuperscript{-1} at the outer edge of the flow region. The results for the horizontal flow velocity for distances larger than 3\,pixels from the sunspot boundary have already been reported in Sect.\,3.1.1 of Paper\,I.\par
During sunspot decay, the sunspot loses the penumbra to become a naked spot. While the penumbra dissolves, the horizontal velocity within the spot decreases (see Figs.\,\ref{fig:inflow_AR11646} and \ref{fig:inflow_AR11841}, both panel (d)) as well as the velocity at the boundary of the spot. At some point, the horizontal velocity within the spot becomes smaller than the velocity in the surrounding flow cell. Although this process shows a different duration for the eight spots, this change of the velocity profile is a common behaviour. The flow profile outside the spots shows a change as well. This was already analysed in Paper\,I, and therefore it is not discussed here in detail.\par
After the penumbra is dissolved, the horizontal velocity profile changes. The flow profiles of the spots show individual characteristics. In general, those are caused by the individual evolution of the spots, for example, the reshaping of their morphological structure (Paper\,I). The location of the spot also influences the measurements because the measurement methods reduce the ability to determine a horizontal flow close to the disc centre. Those cannot be compensated for by taking the heliocentric angle into account. During the analysis, all spots have to cross the meridian at some point. The results obtained on those days of the spots' evolution are treated with caution.\par
Three velocity values which characterise the horizontal velocity profile are determined: 
\begin{description}
\item[1.] The maximum of the absolute horizontal velocity inside the spot, $v^{\mathrm{i}}_{\mathrm{h}}$.
\item[2.] The horizontal velocity at the boundary of the spot, $v^{\mathrm{b}}_{\mathrm{h}}$.
\item[3.] The maximum of the absolute horizontal velocity in the surrounding region of the spot, $v^{\mathrm{s}}_{\mathrm{h}}$.
\end{description}
The position of the three values is exemplarily shown in Fig.\,\ref{fig:imd_maps} for AR11841 for the first day of the analysis of the AR (top panels) and six days later (bottom panels) after the spot has lost its penumbra. Coloured ellipses show the three velocity values $v^{\mathrm{i}}_{\mathrm{h}}$, $v^{\mathrm{b}}_{\mathrm{h}}$ , and $v^{\mathrm{s}}_{\mathrm{h}}$ (blue, red, and green lines, respectively). The sign of the horizontal velocity component indicates the direction of the flow. Here, negative velocities describe a flow in the radial direction towards the spot centre. While determining the absolute values, we keep the information of the sign to determine the direction of the flow in the later analysis. For the following description of the results, absolute velocity values are discussed while distinguishing in- and outflows based on the sign. Therefore, the wording `maximum (horizontal) velocity' describes the maximum of the absolute velocity values.\par
The regions for determining the velocities $v^{\mathrm{i}}_{\mathrm{h}}$ and $v^{\mathrm{s}}_{\mathrm{h}}$ both also include the boundary of the sunspot. This leads to the following statements. (1) If the maximum velocity within the spot, $v^{\mathrm{i}}_{\mathrm{h}}$, equals the velocity at the boundary, $v^{\mathrm{b}}_{\mathrm{h}}$, the horizontal velocity of the flow increases (decreases) with radial distance from the spot centre and no real maximum (minimum) for the velocity within the sunspot exists. (2) If the velocity at the boundary, $v^{\mathrm{b}}_{\mathrm{h}}$ , equals the maximum velocity of the surroundings, $v^{\mathrm{s}}_{\mathrm{h}}$, the velocity continuously decreases with radial distance to the sunspot boundary. Thus, the three values describe the profile of horizontal velocity of the flows within the spot and its surroundings.\par
\begin{figure*}[ht!]
\begin{center}
\includegraphics[width=1.\textwidth]{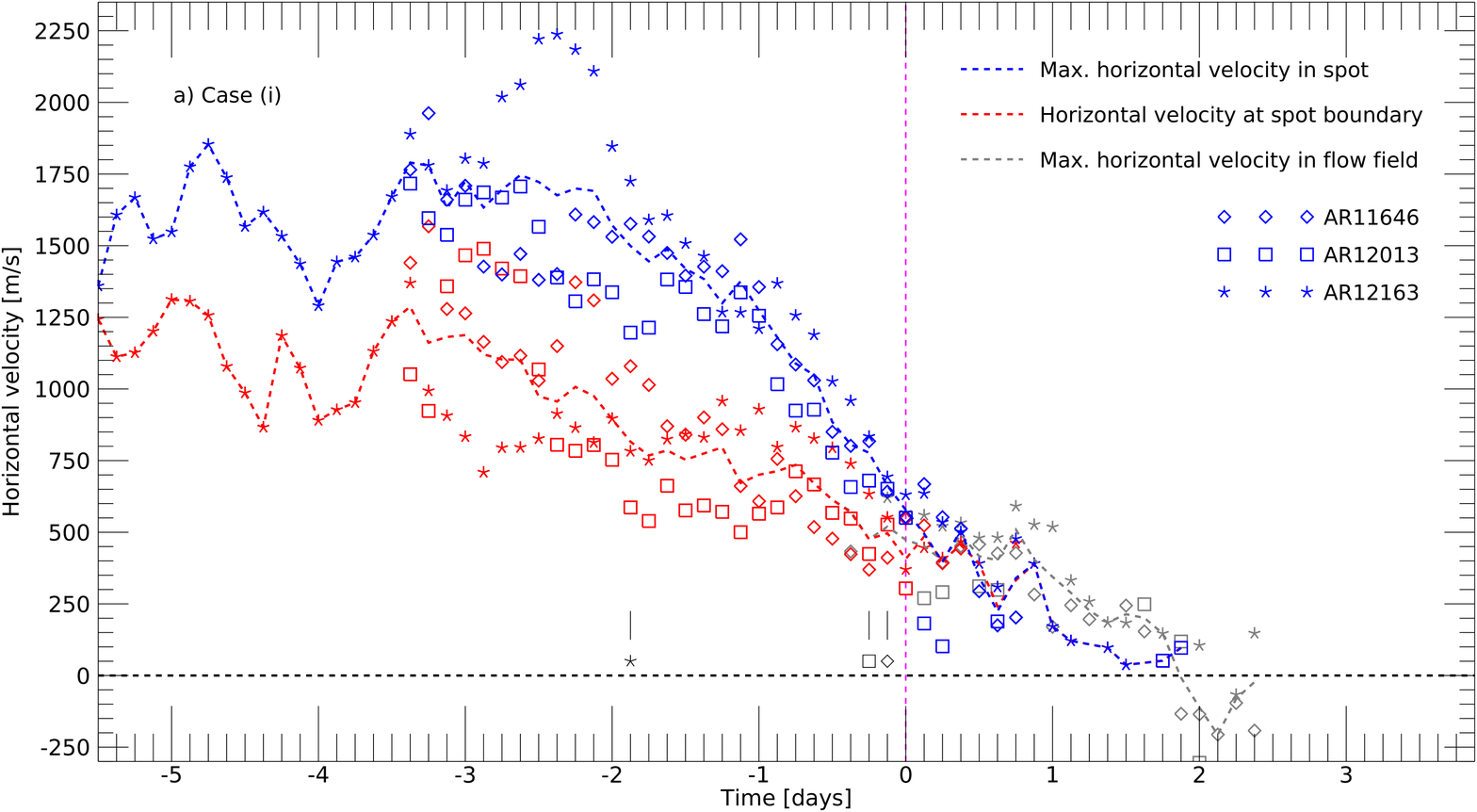}

\includegraphics[width=1.\textwidth]{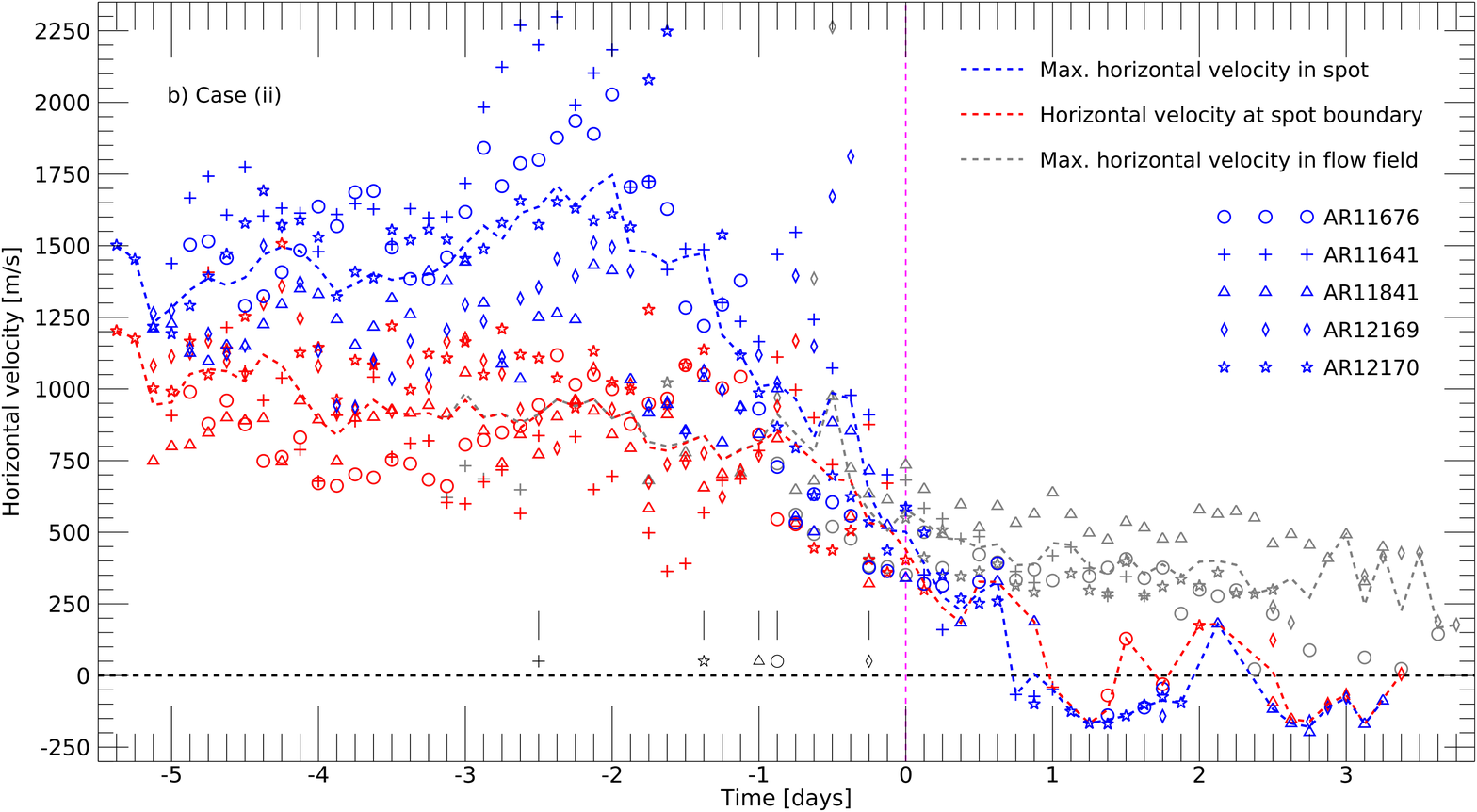}
\end{center}
\caption{Evolution of the maximum horizontal velocity of the flow within the spot boundary (blue), its surrounding region (grey), and the horizontal velocity at the spot boundary (red) for three selected active regions for Case (i) and five active regions for Case (ii). The loss of the penumbra was set as a common point in time, i.e. $t$\,=\,0\,h. Black symbols and short vertical lines indicate the crossing of the meridian of the respective AR within the evolution. \label{fig:flow_profile}}
\end{figure*}

The three different velocity values, $v^{\mathrm{i}}_{\mathrm{h}}$, $v^{\mathrm{b}}_{\mathrm{h}}$ , and $v^{\mathrm{s}}_{\mathrm{h}}$ are shown in Fig.\,\ref{fig:flow_profile} (blue, red, and grey symbols, respectively) for all eight sunspots and at all time steps. The disappearance of the penumbra was set as a common time, that is, $t$\,=\,0\,h (dashed magenta line) for all studied spots. Thus, negative times represent the sunspot with penumbra. For times $t$\,>\,0\,h, the sunspots are declared to be naked spots. The different development of the spots and their stage of evolution when the analysis started leads to a different negative start value $t_0$ for the individual spots. The same holds for the duration of the analysis. The crossing of the meridian of the ARs is indicated by the respective symbol in black and short vertical lines. Besides the individual velocity values for the active regions (symbols), we determined the mean for each time step (dashed profiles).\par
%
If the maximum (absolute) velocity within the spot, $v^{\mathrm{i}}_{\mathrm{h}}$, (blue) is localised at the boundary of the spot, the symbol for the velocity at the spot boundary, $v^{\mathrm{b}}_{\mathrm{h}}$, (red) is not shown. If the maximum velocity in the surroundings of the spot, $v^{\mathrm{s}}_{\mathrm{h}}$, (grey) is equal to the velocity at the spot boundary, $v^{\mathrm{b}}_{\mathrm{h}}$, (red) the former one is not shown.\par
The obtained velocity values, shown in Fig.\,\ref{fig:flow_profile}, represent the description of the flow profiles, as given above. The maximum velocity within the sunspot is largest while the sunspots have a penumbra. The average velocity (blue) does not show any tendency in both cases, until approximately two days before the penumbra has dissolved. Then the horizontal velocity within the sunspots starts to decrease. This is represented by the decrease in the maximum horizontal velocity (blue). The velocity values determined within the spots, $v^{\mathrm{i}}_{\mathrm{h}}$, and those at the spot boundary, $v^{\mathrm{b}}_{\mathrm{h}}$, converge into each other when the penumbra has dissolved. Further in the evolution, the profile of the horizontal velocity shows two different behaviours.\par
The three active regions AR11646, AR12013, and AR12163 show a common behaviour, which in the following is referred to as Case (i). The loss of the penumbra leads to a further decrease in the overall horizontal velocity (see Fig.\,\ref{fig:flow_profile}, panel (a), for t > 0\,h). The two velocities, $v^{\mathrm{i}}_{\mathrm{h}}$ and $v^{\mathrm{b}}_{\mathrm{h}}$ , equalise. Thus, the maximum velocity is no longer localised within the spot, but it is found at the boundary of the naked spot. Within two days, the naked spots disappear. The evolution of the flow profile is exemplarily shown in Fig.\,\ref{fig:inflow_AR11646} for active region AR11646. The penumbra dissolves on the fourth day of the analysis shown in panel d as the turquoise line. The colour-coding is from blue to red over time. The overall maximum horizontal velocity of the studied flow profile is located within the moat region after the penumbra has dissolved. This is represented in the higher values of $v^{\mathrm{s}}_{\mathrm{h}}$ (grey symbols in Fig.\,\ref{fig:flow_profile}, panel (a)) compared to the other two velocity values. The velocity values show a decrease in the horizontal velocity of this flow as well. Thus, the original outflow within the spot, the Evershed flow, and the outflow in its surroundings, the moat flow, become smaller and vanish as these spots dissolve.
\paragraph{Converging flow of naked spots.} The evolution of the flow system of the five other active regions, referred to as Case (ii) in the following, is shown in panel (b)\ of Fig.\,\ref{fig:flow_profile}. Initially, the maximum horizontal velocities within these spots and at the boundary decrease further while the remnant naked spots continue to decay. This agrees with Case (i), representing a decrease in and vanishing of the Evershed flow. Yet, both velocity values, $v^{\mathrm{i}}_{\mathrm{h}}$ and $v^{\mathrm{b}}_{\mathrm{h}}$, become negative within approximately one day after the penumbra has dissolved (see Fig.\,\ref{fig:flow_profile}, panel (b), blue and red symbols, respectively). The negative value represents a change in the flow direction. A flow develops in the opposite direction, that is, towards the centre of the naked spot. In the following, we will refer to it as a `converging inflow'. The converging inflow can be observed until the naked spot has disappeared. Although the flow velocity in the moat region surrounding the spot decreases while the penumbra decays, the flow does not vanish. It remains even after the naked spot has vanished from the solar disc as seen in intensity maps.\par
It should be noted that two factors interfere in the determination of the converging inflow. (1) The number of pixels along an ellipse becomes smaller closer to the centre of the spot. Thus, the number of data points for each fit decreases for smaller ellipses. (2) The smaller velocities lead to a smaller amplitude when the LOS velocity is determined. Therefore, two values are taken into account to determine the validity of the horizontal component of the flow velocity: (i) the standard deviation, which is directly obtained within the fitting procedure, and (ii) the rms value (see Sect.\,\ref{sec:ana}). The rms value is larger than the standard deviation. Only values whose absolute value of the LOS horizontal velocity are larger than the respective rms value are considered to be valid and shown in Fig.\,\ref{fig:flow_profile}. Therefore, some time steps do not show velocity values for individual active regions, even though the spot has not dissolved yet. For example, no values are considered for AR11841 from $t$\,=\,1\,day until $t$\,=\,2\,days (see Fig.\,\ref{fig:flow_profile}, panel (b), blue triangles) while velocity values are considered and shown for later times. For small velocities close to zero, the rms value in general is larger than the absolute value of the LOS velocity. This is mainly the case if the flow changes direction, that is, if the velocity changes from negative to positive. For the overall description of the horizontal velocity profiles, the standard deviation is shown in Fig.\,\ref{fig:inflow_AR11841} as error bars. The extracted horizontal velocity values within the naked spot and the velocity at the boundary of the naked spot resemble each other (see Fig.\,\ref{fig:flow_profile}). The similarity of those values hints to a small radial distance of the position where the values are obtained. Thus, the maximum horizontal velocities of the converging flow within the naked spot are found close to their boundary (see Fig.\,\ref{fig:inflow_AR11841}, panels (f) to (i)).
\section{Discussion}\label{sec:disc}
Here, we expand the study we started in Paper\,I on the evolution of the horizontal flow profile of eight sunspots. We assume an axial symmetrical behaviour of the flows. The obtained velocity profiles for the first days of the analysis, while the sunspots are in a stable stage, are in agreement with previous works \citep[see e.g.][]{Loehner-Boettcher_2013,Verma_2018}. Within the sunspot penumbra, the Evershed flow is measured. Maximum Evershed velocity values between 1.8\,km\,s$^{-1}$ and 2.8\,km\,s\textsuperscript{-1} were reported by \citet{Loehner-Boettcher_2013}. This agrees well with our maximum horizontal velocities, although our range extends down to 1.1\,km\,s\textsuperscript{-1}. The Evershed and the moat flow are well-known phenomena of stable sunspots. A detailed comparison of the characteristics of the moat flow found for these sunspots is provided in Paper\,I. Our findings in that paper are supported by several other studies. In the region closest to the spot the Evershed flow and moat flow mix. Although both flows are not connected, they are difficult to disentangle in Doppler maps. The analysis method assumes roundish spots. However, filamentary structures of the penumbra might extend further out than the determined boundary. Thus, the obtained moat velocities would be contaminated with flow velocities from the penumbra at the proximity closest to the sunspot.\par
As the penumbra decays, the Evershed flow vanishes and, therefore, its influence on the measured velocities in the mixed-zone disappears. Outside the naked spot, characteristics of a normal supergranular velocity profile become visible: the maximum in the velocity profile is found at a larger distance to the spot boundary (see Fig.\,\ref{fig:flow_profile}, grey symbols and Paper\,I). The horizontal flow velocity within and at the boundary of naked spots shows an evolution with two distinguishable scenarios while their spots decay. In a similar manner, the horizontal flow velocity in the surroundings of the spots shows two different evolution scenarios as described in Paper\,I. Thus, the evolution of the flow system connected to spots can be distinguished into two scenarios after the penumbra has dissolved.\par
In the first scenario (Case (i)), the flow system disappears with the naked spot full decay which takes approximately two days (Fig.\,\ref{fig:flow_profile}, panel (a)). For AR11646, a reversal in the flow direction is found for this final stage (see Fig.\,\ref{fig:inflow_AR11646} panel (f), colours range from turquoise to red and Fig.\,\ref{fig:flow_profile}, panel (a), grey symbols for $t$\,>2\,h). We surmise that these horizontal inflow velocities might be caused by the action of the surrounding supergranules, as it was already discussed in Sect.\,3.1.3. of Paper\,I. The action of the surrounding supergranular cells impedes the spot to form a stable flow cell around it. In this scenario, supergranular cells eventually squeeze the sunspot flow cell in the final stage of the decay process. The lack of this flow system surrounding the spot explains the missing inward-outward directed flow (Case (ii)) as long as the spot is present. As described in Paper\,I, the entire cell, including the spot, implodes. The region ends up as an accumulation of magnetic flux in the network (see Paper\,I, Fig.5, panel (d)).\par
Active regions corresponding to Case (ii) tend to decay more slowly. The horizontal (out-)flow velocity within the naked spot and at its boundary decreases. At some point it vanishes and a converging (in)flow develops. Its maximum velocity is localised within the naked spot, but it is still close to its boundary. This is visualised in the bottom panels of Fig.\,\ref{fig:imd_maps}, for AR11841. It is important to note that the blue ellipse in the top panels of Fig.\,\ref{fig:imd_maps} is the location of the maximum horizontal velocity of the Evershed flow, which is directed away from the spot centre, while in the bottom panels of Fig.\,\ref{fig:imd_maps}, it refers to a flow directed to the spot centre. Concurrently, the maximum horizontal outflow velocity in the surroundings of the naked spot does not decrease (see Fig.\,\ref{fig:flow_profile}, panel (b), grey symbols and line). Instead, it stays constant with an average value of 0.4\,km\,s\textsuperscript{-1} (Paper\,I). The evolution of the flow system of the active regions in Case (ii) supports the results obtained in the simulations by \citet{Rempel_2015}. He measured an inflow at the periphery of a simulated naked spot enclosed by an outflow further out. The radial flow velocities of the converging inflow in the surroundings of the simulated spot have values between 1\,km\,s\textsuperscript{-1} and 2\,km\,s\textsuperscript{-1}. These values refer to the radial velocities in the simulation and cannot be compared in a quantitative way with the horizontal velocity amplitudes we determined here (below 0.25\,km\,s\textsuperscript{-1}) from LOS velocities.\par
Also, \citet{Deng_2007} observed an annular region around a naked spot with motions directed towards the spot. This region separates the naked spot from the outflow region. They did not provide exact flow velocities. However, they noted that the flow is weaker than the original moat flow.\par
We find the maximum velocity of the converging flow to be localised within the naked spot. In our study, we assume that the analysed spots are round. An ellipse defines the boundary of the sunspot. The horizontal velocity is determined from such ellipses. The ellipse at the sunspot boundary overestimates the radial position of the actual sunspot boundary. An example of the blue and red contours extending beyond the intensity boundary of the AR11841 naked spot can be seen in the bottom-left panel of Fig.\,\ref{fig:imd_maps}. The same holds for the analyses presented by \citet{Rempel_2015}. This means that the measured converging inflow is actually located at the periphery of the naked spot.\par
It should be noted, that the converging inflows do not appear immediately at the time the spot loses its penumbra. Instead, it takes approximately one day until they can be measured. A delay in the change of the flow pattern with the disappearance of penumbral filaments in intensity maps is also reported by \citet{Murabito_2021} who find a persistent although weaker outflow in the region of the original Evershed flow before the convective granular velocity pattern develops. The flow system around pores and sunspots has been studied in simulations by \citet{Rempel_2011}. From toy models, he found the development of converging flows around pores, which are known from observations \citep[see e.g.][]{Wang_1992}, to be caused by two effects: (i) a geometrical alignment of granules around any cylindrical constraint leading to a mean flow towards it and (ii) the enhanced brightness of granules at the edges of pores causing an asymmetric cooling and driving a converging motion. With the assumption that the moat flow is driven by convection \citep[see e.g.][]{Meyer_1974,Nye_1988}, originates beneath the penumbra, and is directed by the inclined magnetopause of the sunspot, we deduce that the moat flow strongly resembles the flow field of supergranular cells. We also surmise that with the loss of the penumbra, that is, with the attenuation of the inclined magnetopause, the supergranular flow would no longer be enhanced and guided horizontally underneath the visible surface. Instead, with the naked spot resembling a pore in its geometrical structure, a flow system similar to the one found in the simulations by \citet{Rempel_2011} would develop: (i) an outflow spatially separated from the spot and weaker than the moat flow and (ii), a converging annular inflow around the naked spot caused by a rearrangement of the surrounding granules and enhanced radiative cooling. In the later stage of the evolution, the spot dissolves and the supergranular upflow takes over the dynamics of the old spot cell thus becoming a supergranular cell (see Paper\,I).\par
The simulations by \citet{Rempel_2015} predict an inflow-outflow pattern around naked spots. However, our results show that three out of the eight analysed active regions do not develop such a flow system around the naked spots. This is not negligible. Although we did not study the spatial evolution of the magnetic flux of the spots in detail, we can definitely see the difference in the HMI magnetograms for Case (i) and Case (ii) spot cells' magnetic flux by eye. An animation showing the evolution of all active regions in intensity, magnetic field, and velocity is provided as online material in Paper\,I. The spots which develop an inflow-outflow system have close to field-free surroundings when they lose their penumbra. Instead, spots which do not develop the expected flow system become part of the network shortly after their penumbra has dissolved and the host cell has imploded. They are part of a spatially extended magnetic flux accumulation which is only visible in magnetograms. We speculate that this non-cylindrical (according to \citet{Rempel_2011} toy models) spot-and-plage flux system prevents the formation of the Case (ii) inflow-outflow pattern. It should be noted that we have selected mainly isolated H-class sunspots with a well-defined moat region. We surmise that a decay scenario following our Case (i) could be expected for sunspots decaying as part of an active region with several spots or pores in the surroundings, or only partially developed penumbra which often coincides with the lack of a well-developed moat region.
\section{Summary and conclusion}\label{sec:conc}
We have studied horizontal flows in a spot cell, both in and around spots, during sunspot evolution and decay. The spots evolve from a stable stage (fully fledged) into a decay phase (they lose their penumbrae) to eventually vanish from the solar disc. By combining our results with the ones obtained in Paper\,I, we conclude the following evolutionary stages for the flow system of a spot cell:\par
\begin{description}
\item[Stage 1:] With a fully fledged sunspot in the spot cell, the flow system consists of two different flows. In the penumbra, the Evershed flow with strong horizontal velocities dominates. Outside the sunspot, the moat flow takes over.
\item[Stage 2:] While the penumbra decays, the Evershed flow vanishes. At the same time, the velocity of the flow around the spot decreases and is radially displaced in the original moat region.
\item[Stage 3:] After the penumbra has fully dissolved, two different evolutionary scenarios for the spot cell are found. %
In some cases (Case (i) in the Secs.\,\ref{sec:results} and \ref{sec:disc}), the flows vanish with the disappearance of the spot within two days, as seen in intensity maps. The original spot cell implodes under the action of the surrounding supergranular cells causing the remnant (spot) magnetic flux to become part of the network. %
In the other cases (Case (ii) in the Secs.\,\ref{sec:results} and \ref{sec:disc}), a converging flow develops at the direct periphery of the naked spot. Simulations show that this flow is driven by radiative cooling at its periphery. The flow separates the spot from the remaining outflow region. When the spot vanishes, the remnant flow system transforms the original spot cell into a supergranular cell.
\end{description}
We find two scenarios for the evolution of the flow system which is related to a spot and its decay. The spot-flow system is influenced by the evolution of the spot itself and by its interaction with surrounding supergranules. There seems to be a constant interplay between their flows and the flow cell hosted by the spot. A further study focussed on the effect of the surroundings (supergranular flow) in the evolution of the magnetic properties of the spot cell shall provide new insights on the processes behind sunspot and active region decay, such as flux removal, which are still under debate.
\begin{acknowledgements}
The authors thank R. Schlichenmaier, W. Schmidt and M. Rempel for fruitful discussions. We would like to thank the anonymous referee for his thoughtful comments which helped to improve the paper. HS has been funded by the Deutsche Forschungsgemeinschaft, under grant No. RE 3282 and acknowledges financial support from the State Agency for Research of the Spanish MCIU through the "Center of Excellence Severo Ochoa" award to the Instituto de Astrofísica de Andaluc\'ia (SEV-2017-0709). SDO is a mission for NASA’s Living With a Star (LWS) Program. This research has made use of NASA’s Astrophysics Data System.
\end{acknowledgements}
\bibliographystyle{aa}
\bibliography{aa42564-21}
\end{document}